\documentstyle[prl,aps,epsf]{revtex} 

\begin{document}



\title
{ P.-A. Bares and M. Mobilia reply }

\author
{P.-A. Bares and M. Mobilia}

\address
{
Institut de Physique Th\'{e}orique,
\'{E}cole Polytechnique F\'{e}d\'{e}rale de Lausanne,
CH-1015 Lausanne, Switzerland 
}

\date{February, 2000 }

\maketitle

We thank Park et al. for their stimulating comments \cite{Park}.

Notice the following points (in the whole Reply we will adopt the same notation as in \cite{Park,Bares}):

i) It is important to observe  that $H_1$ is not only quartic, but involves  also a
quadratic term ($L\sum_{q}(1-\cos q) a_q a_{-q}$, in their notation) which is essential for the probability conservation. It seems that Park et al. have forgotten this contribution in their computations.

ii) The expression $(2)$ of their comment \cite{Park} is incorrect  and should read $Z[\xi, t_{0}+dt]=\exp\left(\sum_{q>0} \xi_{q}\xi_{-q}[f(q,t_0)-dt(\omega(q)+\omega^{\star} (q)-2\epsilon \sin q))f(q,t_0)+2\epsilon dt \sin q]\right)$, where we use the same notation as Park et al.

iii)Even if we have some doubts on the correctness of the explicit form of their generating functions $Z[\xi,dt]$, we agree that in the general case the latter involves a non-gaussian form. Our original approach follows, however, a different path.

Let us illustrate, in the simplest case, what we meant in writting \cite{Bares}:``Because of the conservation of probability, an analog of the Wick theorem applies and all multipoint correlation functions can be computed''.
To do so we recall that $H=H_0+\frac{\gamma}{L}H_1$, where $H_0=\sum_{q>0} \left[\omega(q) a_q^{\dagger}a_q +\ 
\omega^{\star}(q) a_{-q}^{\dagger}a_{-q} +2\sin{q}
\left(\epsilon'a_q a_{-q} + \epsilon a_{-q}^{\dagger}
a_{q}^{\dagger}\right)\right]+\epsilon L\nonumber $ and 
$H_1=-L\sum_q (1-\cos q)a_q^{\dagger}a_q + 
 \sum_{q_1,q_2,q_3} \cos(q_1-q_2) 
a_{q_1}^{\dagger}a_{q_2}a_{q_3}^{\dagger}a_{q_1+q_3-q_2}\ $, where
 $H_0$ is the {\it free-fermionic} part of $H$.
Now, because of the conservation of probability, we have $\langle \widetilde \chi|H =0$. Morever, $H_{0}$ being itself {\it stochastic}, we also have $\langle \widetilde \chi|H_{0}=\langle \widetilde \chi|H_{1}=0$. 

This implies then,
\begin{eqnarray}
\label{1}
L\sum_q (1-\cos q)\langle \widetilde \chi| a_q^{\dagger}a_q =
 \sum_{q_1,q_2,q_3} \cos(q_1-q_2) \langle\widetilde \chi|  a_{q_1}^{\dagger}a_{q_2}a_{q_3}^{\dagger}a_{q_1+q_3-q_2}
\end{eqnarray}
Because of the translational invariance ($q\neq 0$), we have $\langle \widetilde \chi|a_{q}^{\dagger}a_{q'} e^{-Ht}|\rho\rangle= \langle a_q^{\dagger}a_{q'}\rangle (t) \delta_{q,q'}= g(q,t)$. We also define
$F_{n}(t)\equiv \int_{-\pi}^{\pi}\frac{dq}{2\pi} g(q,t) \cos nq$.

It follows from (\ref{1}), that 
\begin{eqnarray}
\label{2}
 \frac{1}{L^2}\sum_{q_1,q_2,q_3} \cos(q_1-q_2) \langle a_{q_1}^{\dagger}a_{q_2}a_{q_3}^{\dagger}a_{q_1+q_3-q_2}\rangle (t)=\frac{1}{L}\sum_{q}(1-\cos{q})\langle a_q^{\dagger}a_{q}\rangle (t)=F_0(t)-F_1(t)
\end{eqnarray}
Now, the translational invariance implies that
\begin{eqnarray}
\label{3}
 \langle n_m n_{m+1}\rangle(t)=\frac{1}{L^2}\sum_{q_1,q_2,q_3} \cos(q_1-q_2) \langle a_{q_1}^{\dagger}a_{q_2}a_{q_3}^{\dagger}a_{q_1+q_3-q_2}\rangle (t)
\end{eqnarray}
We then obtain the following expression for ${\cal C}_1(t)$ (with the same notation as in \cite{Bares}):
\begin{eqnarray}
\label{4}
 {\cal C}_1(t)= \langle n_m n_{m+1}\rangle(t)-(\rho(t))^2= F_{0}(t)-F_{1}(t)-(F_{0}(t))^2
\end{eqnarray}
Such an expression, which only comes from the probability conservation, can also be obtained via an ``{\it \`a la Wick}'' factorization of the correlation functions, as given by Eq.$(9)$ of Ref.\cite{Bares}, and this independently of the form of $Z[\xi,t]$.

Let us close this Reply with the following remarks:

iv) Our approach provides, in the model considered, results which are certainly beyond traditional Hartree-Fock theory. In fact the latter predicts decay of the density as $t^{-1}$, for the model under consideration (when $\epsilon=0$), instead of the correct behavior, $\rho(t)\sim (4\pi(h+h')t)^{-1/2}-\gamma(\pi\epsilon' (h+h')t)^{-1}$  \cite{Krebsetal}, which we can reproduce \cite{Bares}  (it has to been emphasized that we obtain not only the leading term of $\rho(t)$ but also the subdominant one).

v) $\forall \gamma$, we are able to provide the exact and non-trivial steady-states \cite{Mendoca}.\\

The support of Swiss National Fonds is acknowledged.


\begin{thebibliography}{99}
%
\bibitem{Park}
 S.-C. Park, J.-M. Park, D. Kim, Comment submitted to Phys. Rev. Lett.
%
\bibitem{Bares}
 P.-A. Bares, M. Mobilia,  Phys. Rev. Lett.\, {\bf 83}, 5214,\,\, (1999)

\bibitem{Krebsetal}
K. Krebs,\,M.P. Pfannm\"uller,\, H. Simon and 
B. Wehefritz, J. Stat. Phys. {\bf 78}, 1471 (1995). 
%
\bibitem{Mendoca}
J.R.G. de Mendo\c{c}a and M.J. de
Oliveira, J. Stat. Phys. {\bf 92}, 651 (1998)
\end{thebibliography}
\end{document}